\documentclass[aps,prb,showpacs,letterpaper]{revtex4}
\usepackage{graphicx}
\usepackage{comment}
\begin{document}
\title{Fictitious Cavity Approach to the Casimir Effect}
\author{W. Luis Moch\'an}
\affiliation{Instituto de Ciencias F\'isicas, Universidad Nacional
  Aut\'onoma de M\'exico, Avenida Universidad S/N, Cuernavaca, Morelos
  62210, M\'exico.}
\author{Carlos Villarreal}
\affiliation{Instituto de F\'{\i}sica, Universidad Nacional Aut\'onoma
de M\'exico, Apartado Postal 20-364, 01000 Distrito Federal, M\'exico.}
\begin{abstract}
  We obtain expressions for the Casimir energy and force following an
  approach which may be applied to cavities made up of arbitrary
  materials. In the case of planar cavities we obtain the well known
  Lifshitz formula. The approach is easily generalizable to other
  geometries. 
\end{abstract}
\maketitle

\section{Introduction}

Half a century ago, Casimir \cite{casimir} predicted that the quantum
fluctuations of the electromagnetic field within a planar cavity would
produce an attractive macroscopic force on its boundaries. His
prediction was based on the properties of the field when confined by
perfectly reflecting mirrors. It is only recently that experimental
studies have attained the necessary accuracy to test in detail the
theoretical predictions;\cite{lamoreaux} the Casimir effect has now
been measured\cite{mohideen,capasso,onofrio,decca} with uncertainties
as small as 1\% 
and at distances\cite{PRA69} down to
$\approx 60 $nm. Therefore, theories of the Casimir force
that account for the properties of realistic cavities have become
indispensable. In 1956 Lifshitz proposed
a macroscopic  theory for two semi-infinite homogeneous dielectric 
slabs\cite{lifshitz} characterized by complex frequency dependent
dielectric functions.
The stress tensor was obtained from the self-correlation of the
fluctuating electromagnetic field whose source consists of fluctuating charge
and
current densities within each slab. Their
autocorrelations are related 
through Kubo's formalism,\cite{kubo57} causality, and the
fluctuation-dissipation theorem, to the complex dielectric 
response. In his calculation, Lifshitz considered only homogeneous and
isotropic media and it was assumed that fluctuating sources at a
given position were completely uncorrelated with sources at nearby
positions so that the results were directly  applicable to
semiinfinite homogeneous flat local media, and did not cope with more
complex systems, such as thin films, layered systems, superlattices, 
photonic crystals and metamaterials.\cite{henkel05} For the same
reason, it seemed incapable of dealing with spatial
dispersion and screening at realistic surfaces.\cite{Ansgar,
  Feibelman,Mochan83}

Several alternatives to the derivation of Lifshitz have been
proposed. Barash and Ginzburg\cite{barash75} determined 
the allowed frequencies $\omega_\ell$ of the cavity modes by solving
Maxwell's homogeneous equations and 
imposing planar boundary conditions. The energy could be obtained from the
resulting density of states if there were no
dissipation. Nevertheless, dissipation yields complex frequencies and
the interpretation of the solutions as normal modes looses meaning.
Barash and Ginzburg overcame this problem by introducing an
auxiliary non-dissipative system. The use of an auxiliary system was
further developed by Kupiszewska\cite{kupiszewska92} in a 1D
calculation in which the problem of quantizing a dissipative 
system is attacked by accounting both for the dynamics of the vacuum
modes and of the atomic dipoles to which they  couple and which make up
the material, together with a thermal
reservoir in which the atomic radiators dissipate the absorbed
energy. A disadvantage of this approach is that it 
requires an explicit microscopic model for the walls of the cavity and
for the thermal bath, thus appearing to restrict its generality.
A similar approach\cite{matloob} was based on a 
Green's function method and Kubo's theorem. In both cases the stress
tensor is obtained from the vacuum modes with an explicit dependence
on the dielectric response $\epsilon(\omega)$. An alternative
treatment of the Casimir force 
was introduced by Jaekel 
and Reynaud\cite{jaekel91}, who calculated the radiation pressure
within a cavity bordered by partially transmitting but lossless
mirrors. Each mirror was replaced by an infinitesimally thin scatterer
characterized by a unitary, energy conserving scattering matrix. Their
calculation was later generalized to the case of 
lossy optical cavities\cite{genet03} by complementing 
the cavity modes with noise modes in such a way that the total
scattering matrix was unitary. The scattering matrix corresponding to
the cavity modes was then obtained through the optical theorem.

Moch\'an \textit{et al.}\cite{mochan05,esquivel05} have obtained an
expression for the 
Casimir force using both the scattering approach and a dissipationless
ancillary system. They have argued that in thermal equilibrium, all
of the properties of the radiation field within a cavity are
completely determined by the optical reflection
amplitudes of the walls. Thus, the Casimir force may be obtained from
the stress tensor of any system whose reflection
amplitudes  are  identical to those of the real
system. A dissipationless fictitious system with those properties was
conceived: It had infinitely thin walls characterized by a unitary
scattering matrix whose elements corresponding to the optical reflection
amplitudes from within the cavity were chosen to be identical to those
of the real system. The transmission amplitudes were chosen in
such a way that the energy that was not reflected was transmitted
without loss to the vacuum outside of the cavity. This permitted a
full quantum mechanical calculation of the fields,  even when the real
system is 
dissipative. The field modes were quantized and counted
by adding perfect mirrors far away from the walls of the real
cavity. These quantizing mirrors produce a field that mimics the incoherent 
radiation back into the cavity that is responsible for maintaining a
detailed balance and thus the thermodynamic equilibrium in the case of
lossy or dissipative real mirrors. The field that enters the cavity
after being reflected by the quantizing mirrors has a very large,
frequency dependent phase that becomes infinitely large as the mirrors
are moved infinitely far away.

The main result from the work mentioned above is that if Lifshitz
formula is written 
in terms of the reflection coefficients of the walls of the cavity, or
equivalently, in terms of their exact surface
impedance\cite{halevi,stratton}, it becomes applicable to any system
with 
translational invariance along the surfaces and isotropy around their
normal and not only to semiinfinite, homogeneous, local mirrors. Thus,
it may be employed to calculate the 
Casimir force between semiinfinite or finite, homogeneous or layered,
local or spatially dispersive, transparent or opaque, finite or
semi-infinite systems. Through a simple substitution of
the appropriate optical coefficients, the formalism has allowed the
calculation of the Casimir force between photonic
structures,\cite{esquivel01,villarreal02} non-local excitonic 
semiconductors,\cite{deLaLuz04} non-local
plasmon-supporting metals with sharp boundaries,\cite{esquivel05,
esquivel06}, 
and between realistic spatially dispersive metals with a smooth
self-consistent electronic density profile.\cite{mochan05,contreras05}
The relative simplicity of the formalism has allowed its
generalization to non-isotropic systems and the calculation of Casimir
torques.\cite{torres06} With a few modifications, it has also been
employed for the calculation 
of other macroscopic forces, such as those due to electronic tunneling
across an insulating gap  separating two conductors.\cite{procopio06}

Nevertheless, there was a shortcoming in the derivation of the
Casimir force presented in Refs. [\onlinecite{mochan05,esquivel05}] as it was
uncritically assumed 
that a unitary, energy-conserving scattering matrix could be built
through a proper choice of transmission coefficients. Somewhat
surprisingly, it turns out to be
impossible to find a unitary scattering 
matrix for evanescent waves, i.e., for $Q>\omega/c$, where $\vec Q$ is
projection of the wavevector parallel to the cavity walls, $\omega$ is the
frequency and $c$ the speed of light, as a single transmitted
evanescent wave is unable to transport 
energy away from the surface while an incident and a
reflected evanescent wave do transport energy from the cavity towards
the surface of lossy and dissipative systems.\cite{guille} 
Although the contribution of evanescent waves to the Casimir
force could be obtained as an analytic continuation from the region of
propagating waves, it is not obvious {\em a priori} that this
extrapolation would yield the correct result.

Recently,\cite{mochan06} the problem of energy transport in the
evanescent region was dealt with by modifying the fictitious system
introduced in Refs. [\onlinecite{mochan05,esquivel05}] in such a way
that evanescent 
waves within the cavity couple to propagating waves outside the
cavity. This was accomplished by filling completely the region beyond the
infinitesimally thin mirrors with a dispersionless and dissipationless
fictitious dielectric with a large permittivity
$\epsilon_f$. In this way, the fictitious light cone $Q\le
\sqrt\epsilon_f\omega/c$ extends beyond the light cone $Q\le\omega/c$
of the vacuum cavity and in the limit $\epsilon_f\to\infty$ all of the
evanescent waves within the cavity would be able to couple to propagating waves
within the fictitious region. Lifshitz formula was thus proven to be
valid both within and without the light cone.\cite{mochan06}

One drawback of the calculation presented in \onlinecite{mochan06} is
its use of an extremely unrealistic dielectric, with a suspiciously
large, real, frequency independent dielectric constant
$\epsilon\to\infty$. As the Casimir force ought to be determined by the
reflection coefficients of the real mirrors,\cite{esquivel05,mochan05}
all of the details of the fictitious system beyond the walls of the
cavity ought to be superfluous after their contribution to the
reflection amplitude has been accounted for, and it should be possible to setup
the calculation without the need of specifying them.

The purpose of the present paper is to develop yet another derivation of
the Casimir force within cavities with walls made up of arbitrary
materials characterized only by their optical reflection
amplitudes. As in Refs. [\onlinecite{esquivel05,mochan05,mochan06}],
we introduce a dissipationless fictitious system with no degrees of freedom
beyond those of the electromagnetic field and with a cavity whose
walls have the same optical coefficients as the real system. However,
unlike the calculations above, we avoid giving any detail of the
fictitious system beyond the reasonable fact that it should be
consistent with detailed balance so that thermodynamic equilibrium
is satisfied, i.e., on the average, for each photon that is not
coherently reflected at a cavity wall and is therefore either absorbed
or transmitted beyond the system, an identical photon has to be
incoherently injected back into the cavity with no phase relation to
the lost photon. We believe that this derivation of the Casimir force
is quite simple and that it can be readily generalized to
other geometries, allowing the calculation of the dispersion forces in
cavities of varied shapes whose walls are made up of realistic
materials. 

The structure of the paper is the following: First we review briefly
the model employed in Ref. [\onlinecite{mochan06}]. In Sec. \ref{energycons}
we study the scattering matrix of a fictitious interface between vacuum and a
dispersionless dielectric with an additional infinitesimally thin
scatterer that forces the optical coefficients to be the same as those
of the real system and in Sec. \ref{total} we study the reflection
amplitudes for both propagating and evanescent waves after adding
quantizing mirrors. Then, in Sec. \ref{delay} we eliminate the
superfluous details from the calculation, keeping only the delay
$T\to\infty$ before injecting back into the cavity the photons it looses
through absorption or transmission in order to restore equilibrium. In
Sec. \ref{snormal} we obtain the electromagnetic normal modes of the
cavity and the contributions of the cavity walls to the density of
states which we employ in Sec. \ref{thermo} to obtain their
contribution to the thermodynamic properties. Finally, section
\ref{conclusions} is devoted to conclusions.

\section{Energy flow}\label{energycons}

In Ref. \onlinecite{mochan06}  a fictitious system was introduced,
consisting of a vacuum cavity bordered by two infinitesimally thin sheets
followed by dispersionless dielectric slabs which are terminated by
perfectly reflecting mirrors. It was argued that in equilibrium the
electromagnetic field within the real cavity ought to coincide with
the electromagnetic field within the fictitious cavity. 
The reflection amplitude of the
infinitesimal sheet together with the dielectric was chosen to
coincide with the reflection amplitude of the mirrors that make up the
real cavity. The perfect mirrors were incorporated in order to
quantize the normal modes and in order to inject back into the cavity
any radiation that is not coherently reflected at the surface,
guaranteeing thermodynamic equilibrium. The re-injected radiation
acquires a large phase as it travels across the wide dielectric slab,
thus mimicking the incoherent re-radiation of photons that are lost
through transmission or through absorption at the walls of the real
cavity. 

\begin{figure}
  \includegraphics{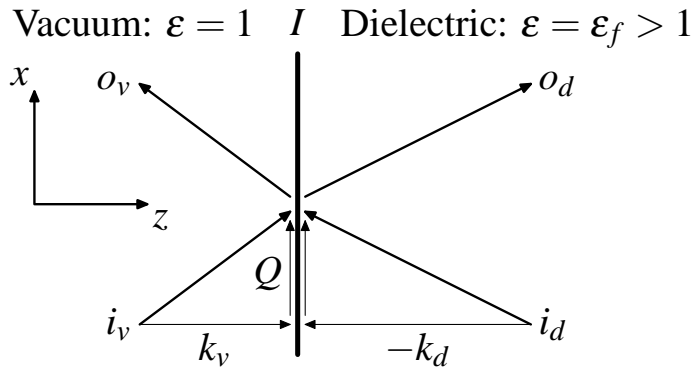}
  \caption{\label{finterface} Infinitely thin scattering interface ($I$)
  between vacuum and a dispersionless dielectric with permittivity
  $\epsilon_f$. Incoming ($i$) and outgoing ($o$) waves within vacuum
  ($v$) and the dielectric ($d$) are illustrated, as well as the
  projection  of their wavevectors parallel ($Q$) and normal ($k_v$
  and $k_d$) to the surface. The choice of coordinate axes is also shown.} 
\end{figure}

Consider one of the mirrors of that fictitious cavity, as illustrated
in Fig. \ref{finterface}, 
consisting of an infinitely thin reflector between vacuum and a
dispersionless dielectric. 
We first assume that all the waves in the figure are propagating,
i.e., $k_v$ and $k_d$ are real, then the energy flux along the normal,
taken to be the $z$ direction, is 
\begin{equation}\label{S1}
  S_{vz}=\frac{c^2}{8\pi\omega}k_v(|i_v|^2-|o_v|^2),\quad
  S_{dz}=\frac{c^2}{8\pi\omega}k_d(|o_d|^2-|i_d|^2)
\end{equation}
 in the vacuum and the
dielectric side 
respectively, where $i_\alpha$ and $o_\alpha$ are the amplitudes of
the incoming and outgoing electric fields within the $\alpha=v,d$
region. Energy conservation requires that
\begin{equation}\label{cons}
  k_v|o_v|^2+k_d|o_d|^2=k_v|i_v|^2+k_d|i_d|^2
\end{equation}
for arbitrary values of
the incoming amplitudes $i_v$ and $i_d$. If we define a scattering
matrix $\mathbf S$ through
\begin{equation}\label{defS}
  \left(\begin{array}{c} \sqrt k_v o_v\\\sqrt k_d o_d\end{array}\right)
    = \mathbf S
    \left(\begin{array}{c} \sqrt k_v i_v\\\sqrt k_d i_d\end{array}\right)
\end{equation}
then $\mathbf S$ should be unitary in the usual sense, i.e.,
$\mathbf S^\dag \mathbf S = \mathbf S \mathbf S^\dag = \mathbf 1$,
where $\mathbf 1$ is the unit matrix. We identify the components of
$\mathbf S$ as
\begin{equation}\label{Sprop}
  \mathbf S = \left (
  \begin{array}{cc} r_v&t_d\sqrt{k_v/k_d}\\t_v\sqrt{k_d/k_v}& r_d
  \end{array}
  \right),
\end{equation}
where $r_\alpha$ and $t_\alpha$ are the reflection and transmission
amplitudes corresponding to incidence on the interface from medium
$\alpha$. Unitarity then yields the relations
\begin{equation}\label{rel1}
  |r_v|^2+\frac{k_d}{k_v} |t_v|^2=1,\quad
  |r_d|^2+\frac{k_v}{k_d} |t_d|^2=1,\quad
  r_v^* t_d\sqrt{\frac{k_v}{k_d}}+ r_d t_v^*\sqrt{\frac{k_d}{k_v}}
  =0,
\end{equation}
which imply
\begin{equation}\label{rel2}
  R_v=R_d,\quad T_v=T_d,\quad R_v+T_v=R_d+T_d=1,
\end{equation}
where we identify as usual the reflectance ($R_v=|r_v|^2$,
$R_d=|r_d|^2$) and transmittance ($T_v=(k_d/k_v)|t_v|^2$,
$T_d=(k_v/k_d)|t_d|^2$) and we denote by $(\ldots)^*$ the
complex conjugate of any quantity ($\ldots$) . Of course,
Eqs. (\ref{rel2}) are consistent 
with Fresnel relations \cite{hecht}. However, we must remark that
$r_\alpha$ and $t_\alpha$ above are not given by the Fresnel relations
due to the presence of an additional scatterer at the interface which
forces the reflection amplitude from the vacuum side to agree with the
reflection amplitude of a boundary of the real
cavity.\cite{esquivel05,mochan05,mochan06}

In the case of evanescent waves, i.e., when $Q>\omega/c$ the normal
component of the wavevector $k_v=i\kappa_v$ is imaginary and the
energy flux is not given by 
Eq. (\ref{S1}) but by
\begin{equation}\label{S2}
S_{vz}=2 \frac{c^2}{8\pi\omega}\kappa_v (i_v^*o_v)''.
\end{equation}
instead of Eq. (\ref{S1}), where we denote by $(\ldots)'$ and
$(\ldots)''$, or equivalently, by $\mathrm{Re}(\ldots)$ and
$\mathrm{Im}(\ldots)$ the real and imaginary parts of any quantity
$(\ldots)$. 
If $Q$ were so large
$Q>\sqrt{\epsilon_f}\omega/c$ so that waves were also evanescent in the
dielectric, we would also have
\begin{equation}\label{S3}
S_{dz}=2 \frac{c^2}{8\pi\omega}\kappa_d (i_d o_d^*)''
\end{equation}
and it would be impossible to satisfy energy conservation $S_{vz}=S_{dz}$
for arbitrary incoming amplitudes. This was the reason for introducing
the fictitious dielectric $\epsilon_f$ and for taking the limit
$\epsilon_f\to\infty$ in Ref. [\onlinecite{mochan06}]. For waves that
are evanescent within vacuum but are propagating within the
dielectric, $S_{dz}$ would be given by Eq. (\ref{S1}) and energy
conservation could be satisfied provided 
\begin{equation}\label{rel3}
  |r_d|^2=1,\quad
|t_v|^2 = 2\frac{\kappa_v}{k_d} r''_v,\quad
t_d=i\frac{k_d}{\kappa_v} t_v^* r_d.
\end{equation}

\section{Total reflection amplitudes}\label{total}

\begin{figure}
  \includegraphics{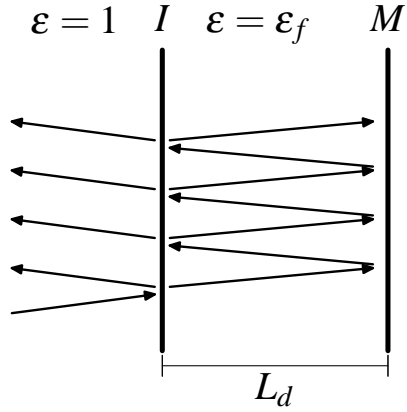}
  \caption{\label{fmultiple}Multiple reflections at a fictitious
    mirror made up of the
  interface shown in Fig. \ref{finterface} with the dielectric
  terminated by a perfect mirror ($M$) after a distance $L_d$.}
\end{figure}

If we terminate the dielectric in Fig. \ref{finterface} a large
distance $L_d$ from the interface and situate a perfect mirror there,
then the total reflection amplitude of the system may be obtained by
adding multiple reflections (see Fig. \ref{fmultiple}),
\begin{equation}\label{multiple}
  r_t=r_v-t_vt_de^{2ik_dL_d}(1-r_d e^{2ik_dL_d} + r_d^2 e^{4ik_d
  L_d}-\ldots)=\frac{r_v+(r_vr_d-t_vt_d)e^{2ik_dL_d}}{1+r_de^{2ik_dL_d}}. 
\end{equation}

If the waves in vacuum are propagating, we may use
Eqs. (\ref{rel1}) to obtain
\begin{equation}\label{rtprop}
  r_t=\frac{r_v+\frac{r_v}{r_d^*}e^{2ik_d L_d}}{1+r_d e^{2ik_dL_d}}.
\end{equation}
Although the reflection amplitude of the real system $|r_v|$ is
typically smaller than 1, it is easy to verify employing
Eqs. (\ref{rel2}) that $|r_t|^2=1$, so
that all of the energy that crosses the interface $I$ eventually
crosses back after some delay. 
On the other hand, for evanescent waves we can use Eqs. (\ref{rel3})
to write
\begin{equation}\label{rteva}
  r_t=2\frac{\mathrm{Re}(r_v + r_v^* r_d e^{2ik_d L_d})}{|1+r_d e^{2ik_d L_d}|^2},
\end{equation}
so that although $r_v$ is complex, $r_t$ is a real quantity. According
to Eq. (\ref{S3}), this means there is no net energy flux towards the
interface so that energy balance would be achieved within the vacuum
cavity also for evanescent waves.

As argued in Refs. [\onlinecite{esquivel05,mochan05,mochan06}], the
properties of the electromagnetic field within a cavity made up of
lossless mirrors such as those described in this section ought to
agree with those within the real cavity. Thus, 
equations equivalent to (\ref{rtprop}) and
(\ref{rteva}) were employed in Ref. [\onlinecite{mochan06}] to
calculate the normal modes of the 
cavity and from them its thermodynamic properties, including the
Casimir force. The purpose of this paper is to discard the unnecessary
and dubious details of the fictitious system, such as the
dispersionless dielectric with a large permittivity.

\section{Delay}\label{delay}

Eqs. (\ref{rtprop}) and (\ref{rteva}) contain elements from the real
system, such as the reflection amplitude $r_v$ of the cavity walls. They
also contain quantities that relate to fictitious quantities,
such as the dielectric constant $\epsilon_f$ of the dielectric, its
width $L_d$ and the reflection amplitude $r_d$ for light impinging on
the interface from the dielectric. To eliminate those quantities from 
the model we first notice that Eqs. (\ref{rtprop}) and (\ref{rteva})
contain terms which are relatively slowly varying functions of the
frequency, such as the reflection amplitudes $r_v$ of the real cavity
walls. On the other hand, they contain extremely fast varying
functions of the frequency such as $e^{2ik_dL_d}$, which are due to the
long time taken by the field to transverse twice the width of the $L_d$
dielectric after entering it across the interface in order to return
after reflection in the perfect mirror.
This long delay is precisely what allows the
fictitious system to replenish those photons that are lost from the
cavity with such a large phase that they mimic the incoherent thermal
photons that would be radiated back into the cavity in the real lossy
system. It is actually the essence of the fictitious cavity
model: In the real cavity photons that are not coherently reflected are
lost through transmission or absorption at the walls; 
the cavity walls would radiate photons incoherently and inject them
into the cavity to sustain thermodynamic equilibrium. In the
fictitious system photons are not lost, but they are delayed a time
$T=2k_dL_d/\omega\to\infty$ before they reach the cavity again, so they are
essentially indistinguishable from incoherently radiated thermal
photons. We can keep the delay $T$ in the model eliminating all other
details by postulating that the reflection amplitude takes the form
\begin{equation}\label{general}
r_t = \frac{r_v+a e^{i\omega T}}{1 + b e^{i\omega T}},
\end{equation}
where $a$ and $b$ are slowly varying functions of the frequency to be
determined. 
The term $r_v$ in the numerator accounts for the coherent
reflection of the real system. The term  $a e^{i\omega T}$  corresponds to
re-radiation by the cavity walls or to thermal photons entering the
system from outside to replenish the cavity losses,
and it includes the long delay $T$ which mimics incoherence. Finally,
it may happen that such a photon is reflected back from  
the interface into the walls and doesn't enter the cavity on its first
attempt. Thus, 
it could be absorbed and re-emitted some time later. The possibility
of multiple re-injection attempts is accounted for by the term $b
e^{i\omega T}$ in the denominator, analogous to the denominator in
Eq. (\ref{multiple}).

In equilibrium all the energy that leaves the cavity has to enter it
again. Thus, for propagating waves the total reflection amplitude must
obey 
\begin{equation}\label{res1}
  |r_t|^2=1,
\end{equation}
which yields
\begin{equation}\label{ab1}
  |r_v|^2+|a|^2+2 \mathrm{Re}(r_v^* a e^{i\omega T})=
  1+|b|^2+2 \mathrm{Re}(b e^{i\omega T})
\end{equation}
after substitution of Eq. (\ref{general}). Separating the slowly and
rapidly varying functions of $\omega$ in the previous equation we
obtain the two equations
\begin{equation}\label{res2}
  b=r_v^* a,\quad |r_v|^2+|a|^2=1+|b|^2
\end{equation}
from which we obtain
\begin{equation}\label{ab3}
  a = e^{i\delta},\quad b=r_v^* e^{i\delta},
\end{equation}
where the real phase $\delta=\delta(\omega)$ is some slowly varying
function of $\omega$. Substituting
Eqs. (\ref{ab3}) in (\ref{general}) we finally obtain
\begin{equation}\label{general1}
  r_t=\frac{r_v+e^{i(\delta+\omega T)}}{1+r_v^* e^{i(\delta+\omega T)}}.
\end{equation}

For evanescent waves, energy conservation requires
\begin{equation}\label{res3}
  r_t''=0
\end{equation}
instead of Eq. (\ref{res1}), which implies
\begin{equation}\label{ab4}
  \mathrm{Im}[r_v+a b^*+(a-r_v^*b) e^{i\omega T}]=0.
\end{equation}
As done above for Eq. (\ref{ab1}) we separate fast and slow varying
functions of $\omega$ to obtain
\begin{equation}\label{res4}
  a=r_v^* b,\quad \mathrm{Im}(r_v+a b^*)=0,
\end{equation}
from which we obtain
\begin{equation}\label{ab5}
  b=e^{i\delta},\quad a=r_v^* e^{i\delta},
\end{equation}
where $\delta$ is again a slowly varying function of the
frequency. Substituting Eqs. (\ref{ab5}) in (\ref{general}) we finally
obtain
\begin{equation}\label{generaleva}
  r_t=\frac{r_v+r_v^* e^{i(\delta + \omega T)}}{1 + e^{i(\delta +
  \omega T)}} = 2 \frac{\mathrm{Re}(r_v+r_v^*e^{i(\delta + \omega
  T)})} {|1 + e^{i(\delta + \omega T)}|^2}.
\end{equation}
Notice that Eqs. (\ref{rtprop}) and (\ref{rteva}) have the form of
Eqs. (\ref{general1}) and (\ref{generaleva}) respectively, where we
identify $e^{i\delta}=r_v/r_d^*$ for propagating waves and
$e^{i\delta}=r_d$ for evanescent waves, as $|r_v/r_d^*|=1$ in the
former case according to (\ref{rel2}) and  $|r_d|=1$ in the latter case 
according to (\ref{rel3}). However, we didn't have to postulate any extraneous
fictitious dielectric slab nor the reflection amplitude $r_d$ from within
it. We will take the limit $T\to\infty$ and we will show that the
unknown phase $\delta$ is irrelevant.

\section{Normal modes}\label{snormal}

\begin{figure}
  \includegraphics{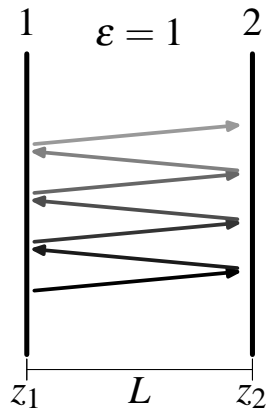}
  \caption{\label{cavity}Cavity made up of two plane, parallel, lossy mirrors
  (1 and 2) with reflection amplitudes $r_1$ and $r_2$ situated a
  distance $L$ apart at $z_1$ and $z_2$. As the field is reflected
  back and forth (arrows) its amplitude diminishes.}
\end{figure}

Given a real dispersive and dissipative system with a coherent
reflection amplitude $r_v$, in the previous section we obtained an
expression for the total reflection amplitude $r_t$ that accounts for
the coherent reflection and mimics the incoherent reflection in
thermodynamic equilibrium through terms that oscillate rapidly as a
function of frequency. Consider now a planar cavity bordered by two
arbitrary material slabs with reflection amplitudes $r_1$ and
$r_2$ (Fig. \ref{cavity}).
For a given $\vec Q$ and $\omega$ the field within the cavity would be
proportional to $A_r e^{ik_vz}+A_l e^{-ik_vz}$, with $A_r$ and $A_l$
the amplitudes of the right and left moving contributions to the
field. Applying boundary conditions at $z_1$ ($A_r e^{i k_v z_1}=r_1
A_l e^{-i k_v z_1}$) and at $z_2$ ($A_l e^{-i k_v z_2}=r_2
A_r e^{-i k_v z_2}$) would yield the usual condition for the {\em
  normal modes} of the cavity,
\begin{equation}\label{normal1}
 D\equiv= 1-r_1 r_2 e^{2ik_v L}=0.
\end{equation}
Due to the losses at the mirrors, we expect that after externally
exciting any of the 
modes obtained from Eq. (\ref{normal1}), it would decay in time
(Fig. \ref{cavity}). Thus, for  
a given real $\vec Q$,  Eq. (\ref{normal1}) would yield a set of
complex frequencies whose imaginary part is related to the finite lifetime of
each mode.  Nevertheless, at thermodynamic equilibrium, the energy of
each mode would be replenished through re-radiation, so that the
equilibrium modes would be stationary instead of decaying. We may
obtain the stationary modes by replacing the reflection amplitudes
$r_1$ and $r_2$ by the corresponding total reflection amplitudes
$r_{1t}$ and $r_{2t}$ obtained through equations such as
(\ref{general1}) and (\ref{generaleva}),
\begin{equation}\label{normalt}
D_t\equiv  1-r_{1t} r_{2t} e^{2ik_v L}=0.
\end{equation}
As $r_{1t}$ and $r_{2t}$ are lossless, the frequencies $\omega_\ell$ that
solve Eq. (\ref{normalt}) for any real value of $\vec Q$ are
necessarily real. Thus, we can straightforwardly apply quantum
mechanical methods to the modes derived from (\ref{normalt}). In
particular, the contribution of the $\ell$-th mode to the energy of the system
is simply $E_\ell=(\langle n_\ell\rangle+1/2)\hbar\omega_\ell$, where
$\langle n_\ell\rangle$ is the average occupation number of the mode.

For propagating waves $|r_{1t}|=|r_2t|=1$ (Eq. (\ref{res1})), so that
Eq. (\ref{normalt}) may be recast as
\begin{equation}\label{arg}
  \arg(r_{1t}r_{2t}e^{2ik_vL}) = 2\pi \ell
\end{equation}
with $\ell$ an integer. Rewriting (\ref{general1}) as
\begin{equation}\label{rit}
  r_{it}=\frac{(1+r_i e^{-i(\delta_i+\omega T_i)})^2}{|1+r_i
  e^{-i(\delta_i+\omega T_i)}|^2} e^{i(\delta_i+\omega T_i)},
\end{equation}
where we introduce the subindex $i=1,2$ to denote each of the two
mirrors, Eq. (\ref{arg}) becomes
\begin{equation}\label{arg1}
  2\arg(1+r_1 e^{-i(\delta_1+\omega T_1)}) + 2\arg(1+r_2
  e^{-i(\delta_2+\omega T_2)}) + \omega T +
  \delta+2k_vL=2\pi \ell,
\end{equation}
where $T=T_1+T_2$ and $\delta=\delta_1+\delta_2$.
Notice that as $\omega$ varies, $1+r_1e^{-i(\delta_1+\omega T_1)}$
moves counterclockwise in the complex plane around a circle centered
at 1 whose radius 
$|r_1|<1$. Thus, it does not encircle the origin and its contribution
to the phase in (\ref{arg1}) is bounded. The same happens with the
second term in (\ref{arg1}). Thus, the eigenfrequencies $\omega_\ell$
never get far away  
from the eigenfrequencies
\begin{equation}\label{wn0}
  \omega_{\ell0}=(2\pi\ell-2k_vL-\delta)/T
\end{equation}
corresponding to {\em vacuum}, for which $r_1=r_2=0$. Notice that as
$T\to\infty$ successive frequencies approach each other and the
density of states diverges.

Consider now a small frequency range $\Omega$ of size $\Delta \omega$ 
centered at a given frequency $\overline \omega$, where
$\Delta\omega$ is much smaller than any characteristic frequency of
the system. Nevertheless, as we will take the limit $T_1, T_2\to\infty$, we
may assume that $\Delta\omega T \gg 1$ so that the number $N(\Omega)$
of normal modes $\omega_\ell$ within $\Omega$ is large $N(\Omega)\gg
1$.  Using Cauchy's argument principle, we obtain
\begin{equation}\label{cauchy}
  N(\Omega)=\frac{1}{2\pi i}\int_\gamma \frac{d}{d\omega}\log f(\omega),
\end{equation}
where $\gamma$ is a clockwise closed path that encircles $\Omega$ and 
\begin{equation}\label{fw}
  f(\omega)=(1+\tilde r_1^*e^{i(\delta_1+\omega T_1)})
  (1+\tilde r_2^*e^{i(\delta_2+\omega T_2)}) - (r_1+
  e^{i(\delta_1+\omega T_1)}) (r_2+ e^{i(\delta_2+\omega T_2)})
  e^{2ik_vL}
\end{equation}
has  within $\gamma$ the same zeroes as
$D_t$ (Eq. (\ref{normalt})) and no poles. Here we introduced linearized and
therefore analytical approximations $\tilde r_1^*$,  $\tilde r_2^*$ to
$r_1^*$,  $r_2^*$, 
taking advantage of the smallness of $\Delta\omega$, so that $f$ is an
analytical function even if $D_t$ is not. 
Choosing $\alpha$ as a path that goes
from $\overline\omega-\Delta\omega/2$ to
$\overline\omega+\Delta\omega/2$ a small distance $\eta\to0$ below the
real axis 
and returns a distance $\eta$ above the real axis, we can rewrite
Eq. (\ref{cauchy}) as
\begin{equation}\label{cauchy1}
    N(\Omega)=\frac{1}{2\pi
    i}\int_{\overline\omega-\Delta\omega/2}^{\overline\omega+\Delta\omega/2}
    d\omega\, \frac{d}{d\omega}\log [f(\omega-i\eta)/f(\omega+i\eta)].
\end{equation}
Approaching the limit $T_1, T_2\to\infty$ before taking the limit
$\eta\to0$, we may assume that $\eta T_1, \eta T_2 \gg 1$ even if
$\eta\ll\Delta\omega$, so that $f(\omega-i\eta)\to e^{i(\delta+\omega
  T)+\eta T}(\tilde r_1^* \tilde r_2^* - e^{2ik_vL})$ and
$f(\omega+i\eta)\to 1-r_1 r_2 
e^{2ik_vL}$. Subtracting the number of modes $N_0(\Omega)$
corresponding to vacuum, which may be obtained from
Eq. (\ref{cauchy1}) by replacing $r_1$ and $r_2$ by zero, we obtain
the contribution $\Delta N(\Omega)$ of the cavity walls to the number
of modes 
\begin{equation}\label{DN}
  \Delta N(\Omega)=-\frac{1}{\pi}
  \int_{\overline\omega-\Delta\omega/2}^{\overline\omega+\Delta\omega/2}
    d\omega \mathrm{Im} \frac{d}{d\omega}\log(1-r_1 r_2 e^{2ik_vL}).
\end{equation}
Notice that all the terms in Eq. (\ref{DN}) are slowly varying and, as
we chose a very small $\Delta\omega$, the integral becomes trivial.
Dividing the result by $\Delta\omega$ we obtain the contribution of the
walls to the density of states
\begin{equation}\label{rho}
  \rho=-\frac{1}{\pi} \mathrm{Im} \frac{d}{d\omega}
  \log(1-r_1r_2e^{2ik_vL}). 
\end{equation}

A similar procedure may be employed for evanescent waves by
substituting Eq. (\ref{generaleva}) instead of (\ref{general1}) in
(\ref{normalt}). The number of modes within the frequency range
$\Omega$ is again given by Eq. (\ref{cauchy}), but choosing
\begin{equation}\label{fw1}
  f(\omega)=(1+e^{i(\delta_1+\omega T_1)})(1+e^{i(\delta_2+\omega
  T_2)}) - (r_1 + \tilde r_1^* e^{i(\delta_1+\omega T_1)}) (r_2 +
  \tilde r_2^* e^{i(\delta_2+\omega T_2)}) e^{-2\kappa_v L},
\end{equation}
where we took $k_v=i\kappa_v$. Now we 
substitute 
$f(\omega-i\eta)\to e^{i(\delta+\omega T)+\eta T}(1-\tilde r_1^*
\tilde r_2^* e^{-2\kappa_v L})$ and
$f(\omega+i\eta)\to 1-r_1r_2e^{-2\kappa_v L}$ in Eq. (\ref{cauchy})
and subtract from the resulting expression the number of modes
corresponding to vacuum.
 The result is again given by Eq. (\ref{DN}), so that the contribution
 of the walls to the density of states for evanescent waves is also
 given by Eq. (\ref{rho}), which is therefore valid  both in the
 propagating  
 ($\omega > Q/c$) and in the evanescent  ($\omega<Q/c$) sectors.

Finally, from Eq. (\ref{rho}) we can obtain an expression for the
contribution of the walls to the average of any real quantity $W_\mu(\vec Q,
\omega)$, namely
\begin{equation}\label{<g>}
  \langle W \rangle = -\mathcal{A}\,\mathrm{Im} \sum_\mu \frac{1}{4\pi^3}
  \int d^2 Q\int   d\omega\, W_\mu(\vec Q,\omega) \frac{d}{d\omega}
  \log (1-\zeta_\mu^{-1}),
\end{equation}
where 
\begin{equation}\label{zeta}
  \zeta_\mu=(r_{\mu 1} r_{\mu 2} e^{2ik_v L})^{-1},
\end{equation}
we incorporated the fact that light has two independent
polarizations $\mu=s,p$ over which we  summed, and we performed the
usual sum over parallel wavevectors $\sum_{\vec Q}\ldots\to
\mathcal{A}/(4\pi^2)\int d^2 Q\ldots$ where $\mathcal A$ is the area of the
mirrors. 

\section{Thermodynamic quantities}\label{thermo}

As simple applications of Eq. (\ref{<g>}) we calculate the ground state
energy 
\begin{equation}\label{U0}
  \mathcal U_0=\langle\hbar\omega/2\rangle=
-\mathcal{A}\, \mathrm{Im} \sum_\mu \frac{\hbar}{8\pi^3}
  \int d^2 Q\int_0^\infty   d\omega\, \omega \frac{d}{d\omega}
  \log (1-\zeta_\mu^{-1}).
\end{equation}
Performing the angular part of the wavevector integration and
integrating by parts over frequency we obtain
\begin{equation}\label{U01}
  \mathcal U_0=
\mathcal{A}\,\mathrm{Im}\frac{\hbar}{4\pi^2}
  \int_0^\infty d Q\, Q\int_0^\infty   d\omega\, \log[(1-\zeta_s^{-1})(1-\zeta_p^{-1})]. 
\end{equation}
Similarly, the internal energy at finite temperature
\begin{equation}\label{U1}
  \mathcal U=\langle \hbar \omega g\rangle = \mathcal{A}\,
  \mathrm{Im}\frac{\hbar}{2\pi^2}
  \int_0^\infty d Q\, Q\int_0^\infty d\omega\, \frac{d}{d\omega}
  (\omega g) 
  \log[(1-\zeta_s^{-1})(1-\zeta_p^{-1})]. 
\end{equation}
where 
\begin{equation}\label{gw}
  g-\frac{1}{2}=\frac{1}{2}
  \coth\left(\frac{\beta\hbar\omega}{2}\right)-\frac{1}{2} 
\end{equation}
is the occupation number of a state with frequency $\omega$ at
temperature $1/k_B\beta$, with $k_B$ Boltzmann's constant. The free
energy 
\begin{equation}\label{F}
  \mathcal F=-\langle \log z \rangle/\beta =\mathcal{A}
  \frac{\hbar}{2\pi^2}\mathrm{Im}\int_0^\infty dQ\, Q \int_0^\infty
  d\omega\, g  
  \log[(1-\zeta_s^{-1})(1-\zeta_p^{-1})], 
\end{equation}
where 
\begin{equation}\label{z}
  z=\frac{1}{2}\mathrm{csch}\left(\frac{\beta\hbar\omega}{2}\right)
\end{equation}
is the partition function of a single mode of frequency
$\omega$. From (\ref{F}) and (\ref{U1}) we may obtain the entropy
\begin{equation}\label{S}
  \mathcal S=\mathcal{A}\frac{\hbar k_B \beta}{2\pi^2}\mathrm{Im}
  \int_0^\infty dQ\, Q \int_0^\infty d\omega\,
  \left(\omega\frac{d}{d\omega}g\right)  
\log[(1-\zeta_s^{-1})(1-\zeta_p^{-1})]. 
\end{equation}
Finally, deriving the free energy with respect to $L$ we may obtain
the Casimir force
\begin{equation}\label{force}
  F=\mathcal{A}\frac{\hbar}{2\pi^2}\mathrm{Re}
  \int_0^\infty dQ\, Q \int_0^\infty d\omega\,
  g k_v
  \left(\frac{1}{\zeta_s-1}+\frac{1}{\zeta_p-1}\right),
\end{equation}
which agrees with Lifshitz formula when written in terms of the
reflection amplitudes. For the actual evaluation of the frequency
integrals in Eqs. (\ref{U01})--(\ref{force}) the integration path can
be conveniently deformed from the real into the imaginary axis
yielding the usual Matsubara summations.

\section{Conclusions}\label{conclusions}

We have obtained expressions for the thermodynamic properties of a
cavity formed by two flat mirrors. We derived our expressions without
making any assumption about the mirrors, which were completely
characterized by their reflection amplitudes $r_1$ and $r_2$. The
mirrors could have been conducting or dielectric, opaque or
transparent, dispersionless or dispersive, lossless or dissipative,
semiinfinite or finite, homogeneous or layered, with abrupt or smooth
boundaries, local or spatially dispersive,  etc.  To derive our
results we postulated a fictitious system which has the same
reflection amplitudes as the real system, but such that any energy
that leaves the cavity eventually comes back but  after a very long
delay which is taken to infinity. Thus, we mimic in a closed system
the incoherent field radiated by the walls of the cavity and the
thermal radiation of the environment which replenishes the energy
lost by the cavity in thermal equilibrium. The requirement of
equilibrium allowed us to find total, lossless reflection amplitudes
for the fictitious system which allowed a full quantum mechanical
description of the system, from which we identified and counted the
normal modes, obtaining an expression for the contribution of the
walls of the cavity to the density of states of the system. With the
density of states, obtaining expressions for all of the thermodynamic
quantities becomes a straightforward task. The expressions we obtained
agree with those found in the literature, although our derivation
shows they are more general than implied by most other derivations. 
We believe that the
procedure is simple enough to be easily extended to non-planar
cavities made up of arbitrary materials.

\acknowledgments

This work was partially supported by DGAPA-UNAM under grant
Nos. IN111306 and IN118605.

\end{document}